\documentclass{aa}
\usepackage{graphicx}
\usepackage{txfonts}
\begin{document}
   \title{El Roque de Los Muchachos Site characteristics. III.}

   \subtitle{Analysis of atmospheric dust and aerosol extinction}

   \author{G. Lombardi \inst{1,2,3}, V. Zitelli \inst{2}, S. Ortolani \inst{4}, M. Pedani \inst{5},
   \and A. Ghedina \inst{5}}

   \offprints{G. Lombardi}

   \institute {Department of Astronomy, University of Bologna, via Ranzani 1, I-40127 Bologna, Italy\\
     \email {gianluca.lombardi@oabo.inaf.it}
     \and
    INAF-Bologna Astronomical Observatory, via Ranzani 1, I-40127 Bologna, Italy\\
   \and
    European Southern Observatory, Casilla 19001, Santiago 19, Chile\\
     \and
    Department of Astronomy, University of Padova, vicolo dell'Osservatorio 2, I-35122 Padova, Italy\\
   \and
    Fundaci\'on Galileo Galilei and Telescopio Nazionale Galileo, P.O. Box 565, E-38700 Santa Cruz
     de La Palma, Tenerife, Spain}

   \date{Received ?; accepted ?}

  \abstract
   {It is known that the Canary Islands are normally affected by dominant winds flowing from north-northeast, that in some meteorological
   conditions, can transport sand from the Sahara desert at high altitude.
   The dust may affect the efficiency of the telescopes and decrease the transparency of the sky.}
   {To maximize the scientific return of the telescopes located at the Observatorio del Roque de Los Muchachos (ORM), we present an analysis of
   the atmospheric dust content and its effects on astronomical observations. Than $B$, $V$ and $I$ dust aerosol astronomical
    extinction are derived.}
   {Using a 5-year series database of data taken from a dust monitor located inside the Telescopio Nazionale Galileo (TNG) dome, we
    computed  mean hourly and daily values of the dust content as measured with a four-channel dust monitor.}
   {We detected particles of 0.3, 0.5, 1.0, and 5.0 $\mu$m. Furthermore, using a power law we derived the content of 10.0 $\mu$m particles. We found a typical local dust
   concentration ranging from $3 \cdot 10^6$ particles per cubic metre at 0.3 $\mu$m, to  $10^3$  at 5.0
   $\mu$m and $10$ at 10.0 $\mu$m, increasing up to 3 orders of magnitudes during the dust storms, with a relatively higher
   increase of 1.0, 5.0, and 10.0 $\mu$m particles. The number of local dust storm events is the same in the local winter
    and summer, but the average background and storm-related increases in the dust concentration in summer are significantly higher than in winter.
   In a uniform approximation, during the dust storms, an average height of the dust layer of 2.5 km above the
   telescope is inferred.}
   {During the sand storms, La Palma Island is affected by an almost uniform layer extending up to 5 km above the sea
   level. The visible extinction is dominated by particles at 1.0, 5.0
   and 10.0 $\mu$m. In agreement with the results from Carlsberg Automatic Meridian Circle (CAMC), we find a typical
   extinction  of about 0.2 mag airmass$^{-1}$ during dust storms.}


   \authorrunning{Lombardi et al.}
   \titlerunning{ORM Site Characteristics. III.}
   \maketitle
%

\section{Introduction}
The Telescopio Nazionale Galileo (TNG) is one of the international
telescopes located at the Observatorio del Roque de Los Muchachos
(ORM) on La Palma Island (Canary Arcipelago).
Due to the proximity of the North African continent, the winds
coming from the central side of the Sahara desert carry
a large amount of dust in suspension that can reach high altitudes and increase the atmospheric dust concentration.\\
Whittet et al. (\cite{whittet87}) pointed that this wind action
removes  approximately $2 \cdot 10^{8}$ tonnes of dust every year
from Sahara desert, and they can be carried out to altitudes up to 6
km. This seasonal action can reach the south of the Canary Islands,
and extended dust clouds often appear above the sky of La Palma,
originating a phenomenon locally named \textit{la Calima}.\\
This natural event is often discuss concerning the quality of
astronomical observations obtained at ORM such when quantify the
contribution of astronomical extinction. Therefore a knowledge of
the amount of the dust and, if possible, of the resulting extinction
is very important for optimizing the performance of both instruments
and telescopes and for obtaining reliable photometric measurements.
Moreover, the knowledge of the grain sizes and their distribution at
the mirror level is fundamental because the spectral index of the
extinction depend on the size of grains. In fact, grains having
dimensions larger than the adopted wavelength produce a neutral
extinction, while sub-micron particles produce an extinction in the same band, with respect to the adopted wavelength.\\
The first detailed study of the impact of Calima on
astronomical observations in La Palma is in Murdin
(\cite{murdin85}), who found a seasonal trend in atmospheric
  transmission, ranging from 90\% of nights classified as good (i.e., nearly
  aerosol-free), between October and March, to 58\% during July, August,
  and September. Another interesting result from Murdin is the atmospheric extinction measured in a site-test
campaign in La Palma. A Murdin summarisation of studies from
other authors shows that 78\% of the clear nights in the months
starting from February to September 1975 and 83\% from December 1974
to November 1975 have extinction less than 0.3 mag
airmass$^{-1}$.\\
To improve the TNG meteorological station, the TNG site group
put in operation a dust monitor. The first preliminary results can
be found in  Porceddu et al. \cite{porceddu02} and Ghedina et al.
\cite{ghedina04}. The primary motivation behind this measurement is
to have a continuous characterisation of the air conditions above
the primary mirror for several reasons. The first one is to compare
dust measurements and atmospheric extinction to know which level of
dust may affect astronomical observations. The second is to have
better maintenance. In fact we know that the deposition of the dust
over the optical surfaces reduces both the optical throughput
(laboratory measurements of the primary mirror reflectivity show a
decrease from 99\% reached in the case of fresh aluminum to 70\% in
the case of a dusty surface) and the signal-to-noise ratio of the
images. The third one is for safety reasons and for coating
lifetimes, when humidity condensation sticks the dust on the mirror.
The knowledge of the level of dust counts helps us to prevent damage
to the optical surfaces of the telescope. Finally, the knowledge of
the dust concentration at different
sizes is also important for the proper design of the sealing of the telescopes
and instruments moving mechanical parts.\\
The first detailed analysis of more than 10 years of meteorological
data obtained using the TNG meteorological station at ORM can be
found in the two recent papers, Lombardi et al. (\cite{lombardi06})
(hereafter Paper I) and Lombardi et al. (\cite{lombardi07})
(hereafter Paper II). Paper I shows a complete analysis of the
vertical temperature gradients and their correlation with the
astronomical seeing. Instead, Paper II shows an analysis of the
correlation between wind and astronomical parameters, as well as the
overall long-term weather conditions at ORM. Differences in the ORM
microclimate demonstrated in a detailed comparison between
synoptical parameters taken at three different locations in the
observatory on a 1000 m spatial scale. Furthermore, ORM was shown to
almost be dominated by high pressure and characterised by an
averaged relative humidity lower than 50\%.\\
Extended dust clouds periodically appear above the ORM complex.
Murdin (\cite{murdin85}) shows that dust collected at the ground
level at ORM  is composed of small round quartz aggregates having
diameters ranging between 10 to 60 $\mu$m diameter. The optical
properties of this dust have been investigated by several authors
(Stickland \cite{stickland87}; Murdin \cite{murdin85}; Jim\'enez et al. \cite{jimenez98}), all founding a flat wavelength dependence of the
extinction; in particular, Murdin points out that
Saharian dust's size is much bigger than the standard photometric $V$ band wavelength.\\
Big plumes of dust during the Calima days are also visible using satellites probes. In Siher et al. (\cite{siher04}) an analysis of
5200 data covering the years 1978-1993
shows a correlation between astronomical extinction and the Total
Ozone Mapping Spectrometer (TOMS) satellite derived extinction as
well as a seasonal trend above the Canaries with a significant
increase in dusty days at the end of July. Siher et al.
(\cite{siher04}) also shows an average of 55 dusty days from the
beginning of June to the beginning of September (about half of the
summer nights). In this paper only the aerosol data corresponding to
a significant absorption were used. The data since 1978 seems to
indicate that the dusty period is drifting during the years. An
independent analysis, based on Carlsberg Automatic Meridian Circle
(CAMC) telescope data, has been published by Guerrero et al. (\cite{guerrero98}).\\

The present paper has the goal of improving the knowledge of the
aerosol-particle-size distribution and the effects on astronomical
observations in La Palma. In particular, we have analysed for the
first time the influence of the dust collected at the level of TNG
primary mirror (M1) as a function of its size to the aim: a) to
understand if local measurements may be taken as good indicators of
the upper conditions in the atmosphere (a dust ground condition has
reproduced a similar situation in the atmosphere?); b) to give an
accurate extinction study selecting among different
sizes of dust; c) to use this information as a primary tool for a fast warning of the unclean environment.\\
The paper is organised as follows. Section \ref{dustmonitor}
describes of the adopted particles counter. Section \ref{seasonal}
describes the adopted procedures to analyse the collected data.
Section \ref{storms} defines and reports number of storm events. A
comparison between local extinction and total visual extinction
obtained from the database
of the CAMC is discussed in Section \ref{extinction}.\\
Using TOMS satellite archive, we have extracted the Aerosol Index
from TOMS/EarthProbe, which provides independent, semi-quantitative
indications of the dust content in the high atmosphere.
Discussions of aerosol and TOMS data are in Section \ref{k_AI},
conclusions can be found in Section \ref{conclusions}.

\section{The dust monitor}\label{dustmonitor}
The selection of air-particles counting equipment is primary
triggered by the particle sizes to be monitored to obtain their
estimated amount. The parameters used for this selection were the
sizes of the dust, the flow rate, and the background noise counts.
Each measured size of the dust corresponds to the number of
available channels of the counter. The sensitivity of the detector
is given by the smallest size of particle that the sensor is able to
detect. The flow rate is the amount of the flux that the sensor can
receive through the sample volume, while the background noise is the
instrument dark current value. We have addressed our choice to an
easy-to-use instrument, hand-held, reliable, and very sensitive.
Moreover, a particle counter based on the optical detection
technique that can count particles above 5.0 $\mu$m diameter was
selected. The reason for such sensitivity is that we expect the
number of dust particles to be less than the total number of
atmospheric particles and we expect them to occur mostly on scales
near 1 $\mu$m with a typical concentration range of a few tens per
cubic centimetre. We therefore chose the particle counter Abacus
TM301 (made by Particle Measuring System, Inc.), which uses a laser
scattering technique for environmental ambient air analysis. Abacus
TM301 is a compact and portable system designed to measure the
purity of close environments, like a clean room, the counter is a
system to measures the density of particles in open air. This system
is based on a filter through which the air is forced to flow.
Different kinds of filters are used sequentially to measure
different sizes of particles centered at 0.3, 0.5, 1.0, and 5.0
$\mu$m according to the estimated dimension of dust and sand of the
Sahara desert as described by Murdin (\cite{murdin85}). A small pump
placed inside the dome sucks the air and, through scattering the
light from a laser diode, measures the number of particles. The four
sizes of particles can be measured in both integrated and
differential modes. Although the counter is mounted inside the TNG
dome and at the level of telescope, a 1 m long silicon pipe through
the wall feeds the pump to external air at 13 m above the ground.
This is a needed precaution because, even if we intend to use the
counter to monitor the external air, it must be kept in a closed
site, safe from adverse weather conditions. Unfortunately, we cannot
distinguish between the type of particles,  but only their size. For
instance, it is not possible to distinguish between water vapour and
dust. For this reason it is necessary to stop the monitoring if
relative humidity increases to the condensation point (typically
$\geq 85\%$). Table \ref{TM} summarises the instrumental basic
performances already reported in Porceddu et al. (\cite{porceddu02})
and  also  reported here for completeness.
     \begin{table}[t]
     \begin{center}
      \caption{Main characteristics of the Abacus TM301 dust monitor.}
         \label{TM}
         \begin{tabular}{r l}
           \hline
           \hline
             \noalign{\smallskip}
            size sensitivity & 0.3 $\mu$m at 50\% counting efficiency\\
            input flow rate  & 0.1 Cubic Foot Minute [CFM]\\
            size channels    & 0.3, 0.5, 1.0, 5.0 $\mu$m\\
            light source     & laser diode ($\lambda = 780$ nm)\\
                        concentration limits & $\geq 10^{6}$ per dm
                        instantaneous\\
            zero counts      & $<1$ per CF\\
                    output           & RS-232\\
                    data storage     & 500 sample, rotating buffer\\
                    sample time      & selectable from 1 sec to 99 minutes\\
                    purge time       & selectable from 1 sec to 99 minutes\\
                    dalay time       & selectable from 1 sec to 99 minutes\\
                        \hline
            \hline
         \end{tabular}
         \end{center}
   \end{table}
\begin{figure*}[t]
\centering
\resizebox{0.8\hsize}{!}{\includegraphics{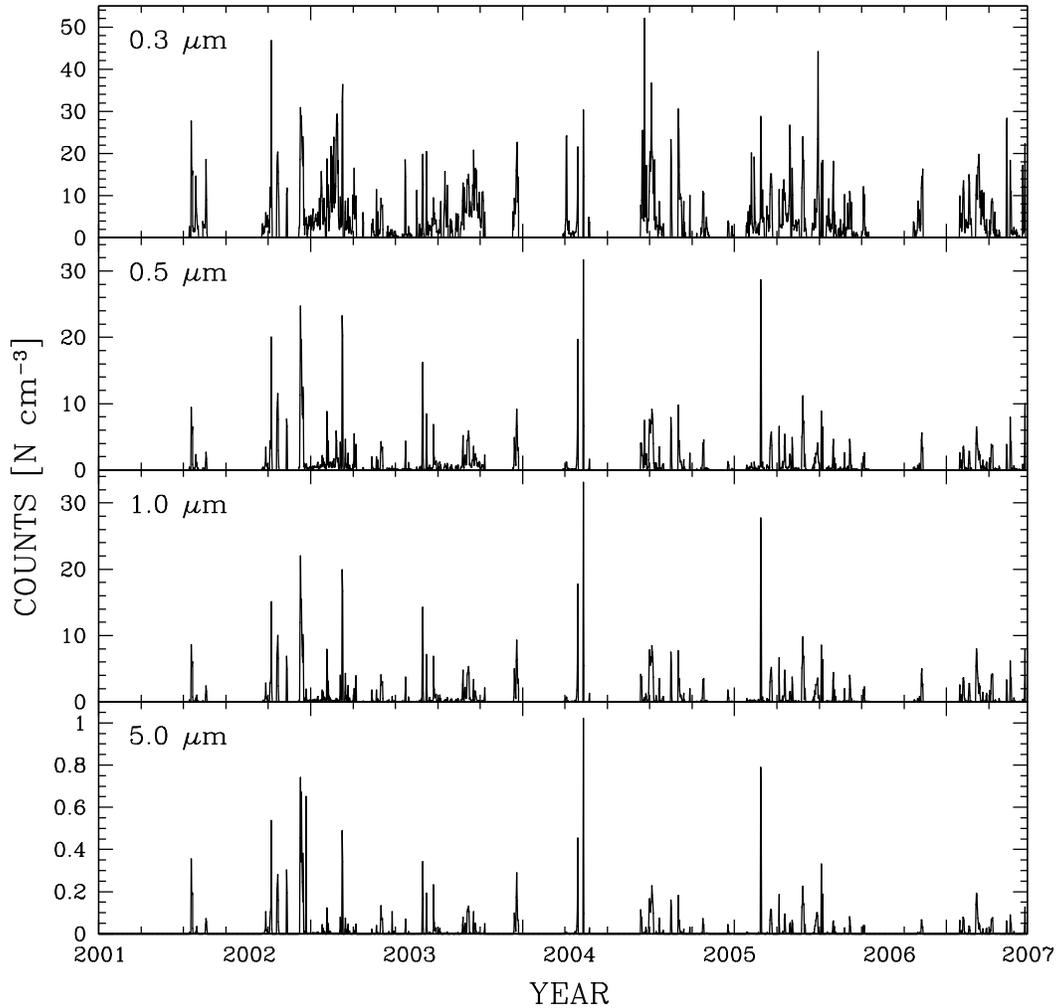}}
\caption{Distribution of the atmospheric particles as a function of
the different size, as measured  by the Abacus TM301 during the
years 2001-2006 (1-2 weeks each month depending on weather
conditions). The gaps correspond to interruptions in the counter
operations or to rejected values due to relative humidity $> 85\%$.}
\label{dust_full_db}
\end{figure*}
\section{Data analysis}\label{seasonal}
The dust monitor measures outdoor dust at the same level as to TNG
M1 to have a monitoring as close as possible the M1 local
conditions. Data coming from Abacus TM301 are considered not as
instantaneous values but as mean values after an integration time of
10 minutes. Before each measurement, a 5-minute long reset is used
to clean the sensor. Abacus TM301 stores  about 500 sets of data in
memory, and when the memory is full, it uses an RS232 serial common
port to transfer data to the PC for further analysis. Up to now the
database collects measurements from August 2001 to December 2006,
and it is used consecutively for 1-2 weeks each month depending to
the weather conditions. In particular the counting is stopped if
relative humidity $> 85\%$. In Figure \ref{dust_full_db} we report
the full dust-count distribution for each particle size as measured
during the  years. The gaps in the figure corresponding to zero
counts are caused by interruptions in the counter operations or by
rejected values due to relative humidity $> 85\%$.\\
To check the presence of a seasonal variation, we split the dust
database in two epochs, defining wintertime  the months from October
until March and summertime the months from April until September. In
this analysis we define the dust storm event by each dust count
having a value a few $\sigma$-levels over the monthly mean values.
The background is evaluated
using the $\sigma$-clipping algorithm also described in Huber (\cite{huber81}) and Patat (\cite{patat03}).\\
The particle counts ($N_{i}$) were assumed to have a Poissonian
distribution. For each month we computed the median value of the
counts ($MED$), then for each different size of dust we calculated
the median absolute deviation ($MAD$) defined as the median of the
distribution
\begin{equation}
\left| N_{i} - MED \right|.
\end{equation}
We set the $\sigma$ parameter to the value $\sigma = 1.48\ MAD$
(according to Huber \cite{huber81}§), where 1.48 is the ratio
between the standard deviation and the $MAD$ under the assumption of
a Gaussian distribution. During the iteration we rejected counts
having
\begin{equation}
\left| N_{i} - MED \right| > K\sigma.
\end{equation}
The procedure for estimating the dust background was iterated twice.
In the first iteration we rejected the points exceeding $\pm
3\sigma$ with respect to the computed monthly median. In the second
iteration, we rejected points exceeding $2\sigma$ and $-3\sigma$ so
that the background distribution is uncontaminated by any other dust
peaks corresponding to dust storms.\\
\begin{figure}[t]
\centering
\includegraphics[width=8cm]{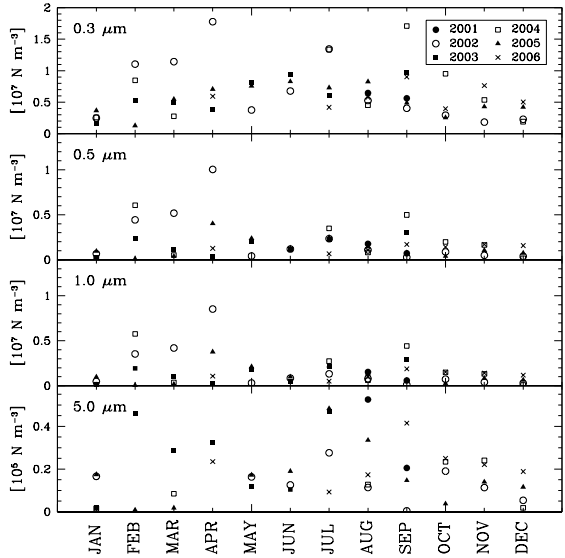}
\caption{Seasonal distribution of the monthly dust at ORM: ($top$) 0.3 $\mu$m dust; ($middle-top$)
 0.5 $\mu$m; ($middle-bottom$) 1.0 $\mu$m dust; ($bottom$) 5.0 $\mu$m dust. Different years are indicated by different symbols.}
\label{monthlydust}
\end{figure}
\begin{table}[b]
     \begin{center}
     \caption[]{Dust background content in [N m$^{-3}$] at ORM in wintertime, summertime, and in the entire annual cycle.}
     \label{median_dust}
     \begin{tabular}{c c c c}
     \hline\hline
 & wintertime & summertime & annual\\
\hline
0.3 $\mu$m & $1.27 \cdot 10^{6}$ & $4.38 \cdot 10^{6}$ & $2.98 \cdot 10^{6}$\\
0.5 $\mu$m & $1.21 \cdot 10^{5}$ & $3.76 \cdot 10^{5}$ & $2.53 \cdot 10^{5}$\\
1.0 $\mu$m & $0.51 \cdot 10^{5}$ & $1.51 \cdot 10^{5}$ & $1.02 \cdot 10^{5}$\\
5.0 $\mu$m & $0.66 \cdot 10^{3}$ & $1.47 \cdot 10^{3}$ & $1.09 \cdot 10^{3}$\\
\hline
\hline
    \end{tabular}
    \end{center}
\end{table}
We have found that, in connection with dust storms, counts of
particles having diameters of 0.3, 0.5, and 1.0 $\mu$m increase by 2
orders of magnitude with respect to the background. Unfortunately
5.0 $\mu$m particles often present a typical fluctuation between 1
and 2 orders of magnitude, so it is difficult to distinguish dust
storms from the typical dust background. For this reason an increase
of 3 orders of magnitudes does allow us to distinguish the 5.0 $\mu$m peaks from the background.\\
Figure \ref{monthlydust} shows the median monthly distribution of
the particles as a function of the month.
It is evident that the level of background depends on size. The
distribution shows the increase in the counts during
February-April and July-September, while  May and June are the months with the lower level of dust.\\
Table \ref{median_dust} shows the median dust content as computed in
the two seasons and in the entire annual cycle. The table clearly
shows that small particles are dominant both in wintertime and in
summertime. It is not surprising because heaviest
particles decay quickly, while lighter particles flow in the atmosphere for longer.\\

\begin{figure}[t]
\centering
\includegraphics[width=8cm]{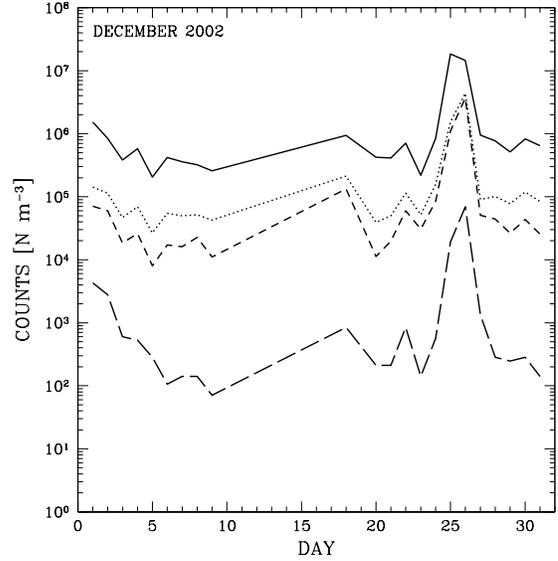}
\caption{Dust storm event of 2002 December 25 and 26: 0.3 $\mu$m
(solid), 0.5 $\mu$m (dots), 1.0 $\mu$m (short-dashes) and 5.0 $\mu$m
(long-dashes).} \label{event}
\end{figure}
\begin{figure}[b]
\centering
\includegraphics[width=8cm]{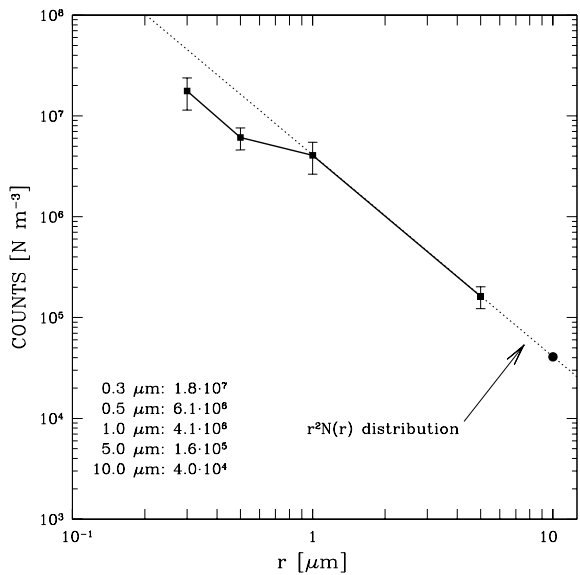}
\caption{Distribution of the median dust counts during dust
storms. Particles $\geq$$1.0$ $\mu$m are treated as if they follow
an $r^{2}N(r)$ power law.} \label{dust_distribution}
\end{figure}

\section{Dust storms}\label{storms}
As mentioned, dust storm is the event where the content of dust,
given in counts per cubic metre [N m$^{-3}$], increases with respect
to the typical dust background content in clear days by 2 orders of
magnitude or more for 0.3, 0.5, and 1.0 $\mu$m dust, and 3 orders of
magnitude or more for 5.0 $\mu$m dust. Figure \ref{event} shows a
more detailed situation of a dust storm that occurred in December
2002. This event persisted for two days, on 2002 December 25 and 26, and it
 shows an increase of several orders of magnitude in the counts.\\
\begin{figure}[t]
\centering
\includegraphics[width=4cm]{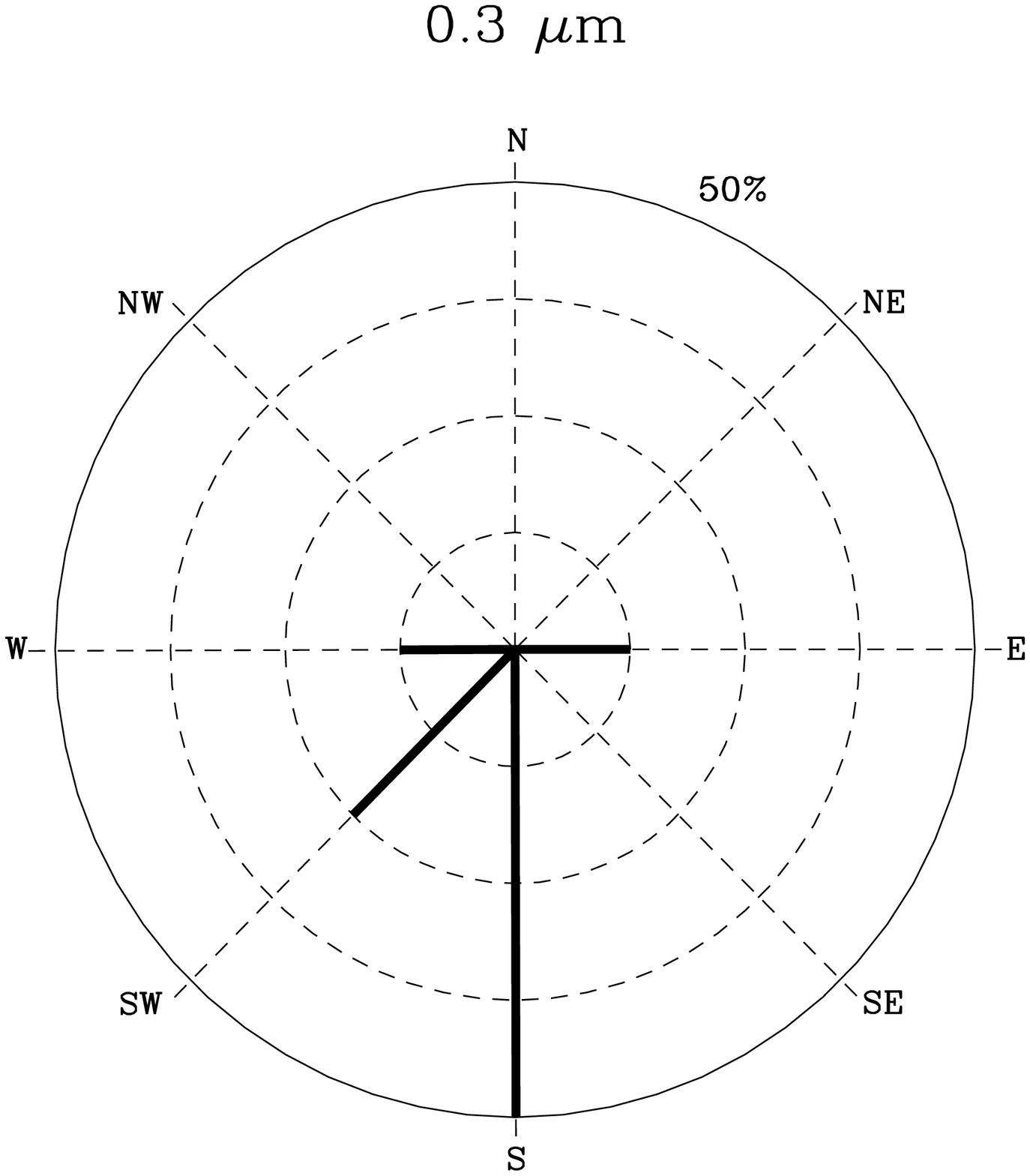}
\includegraphics[width=4cm]{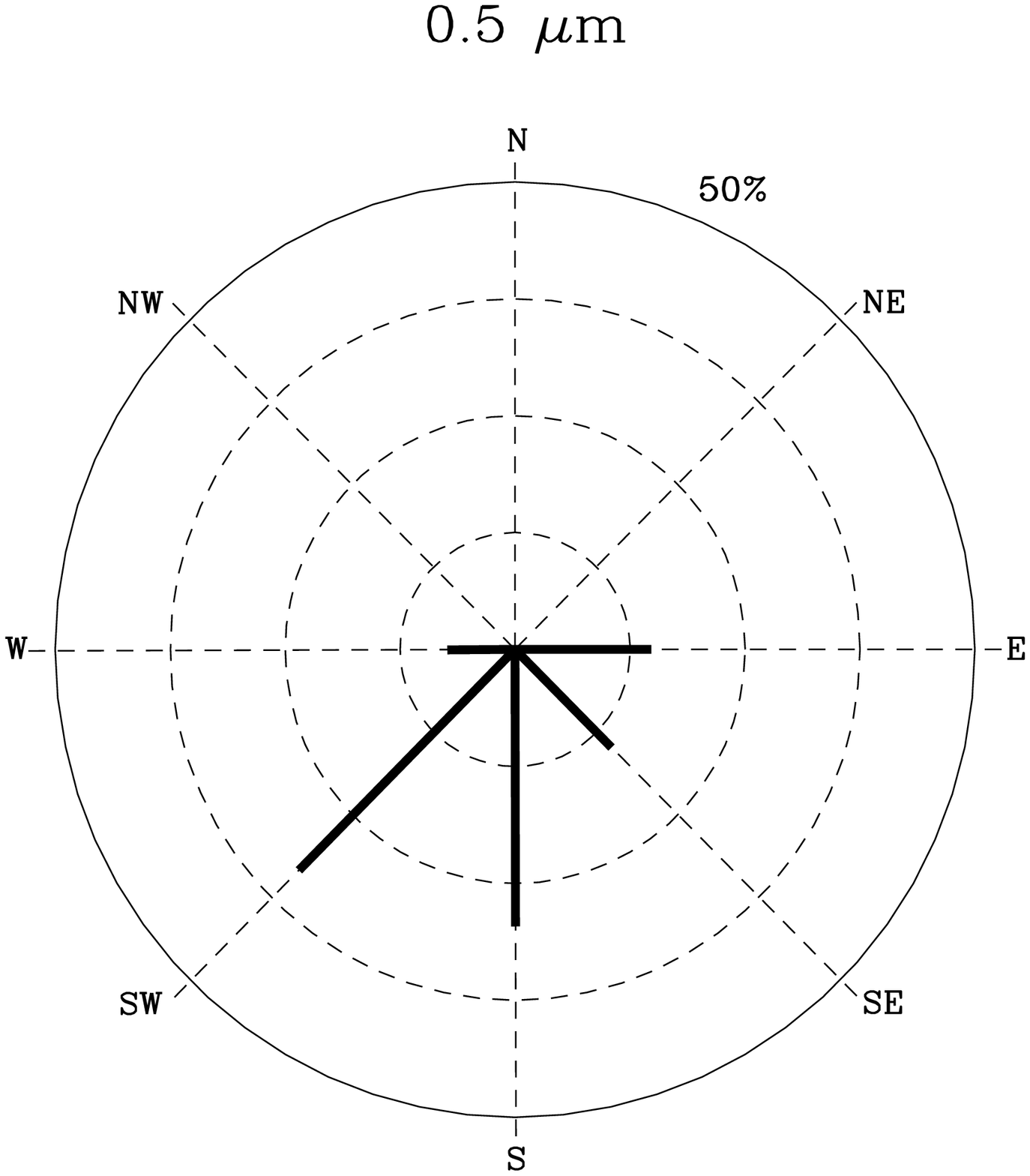}
\includegraphics[width=4cm]{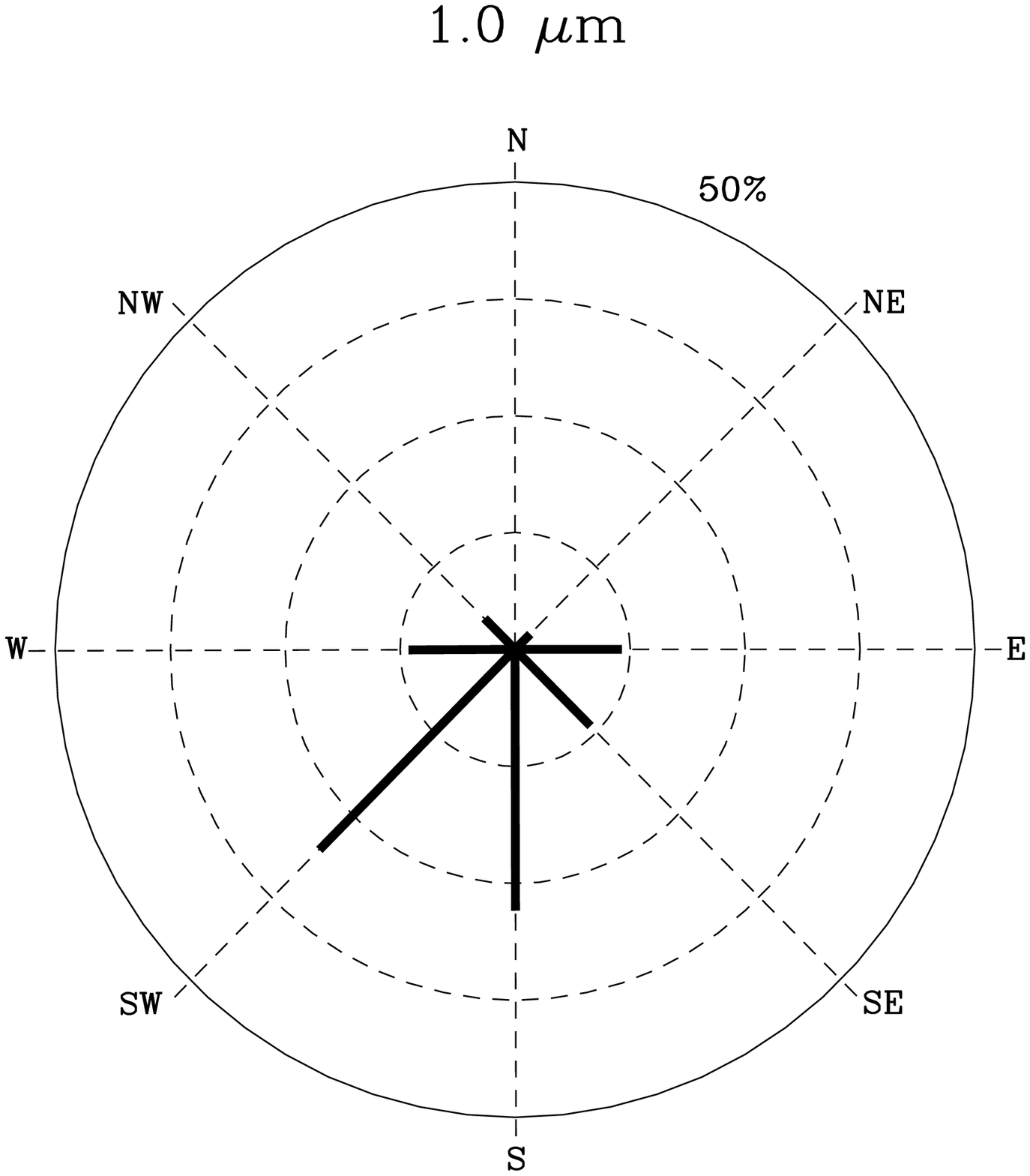}
\includegraphics[width=4cm]{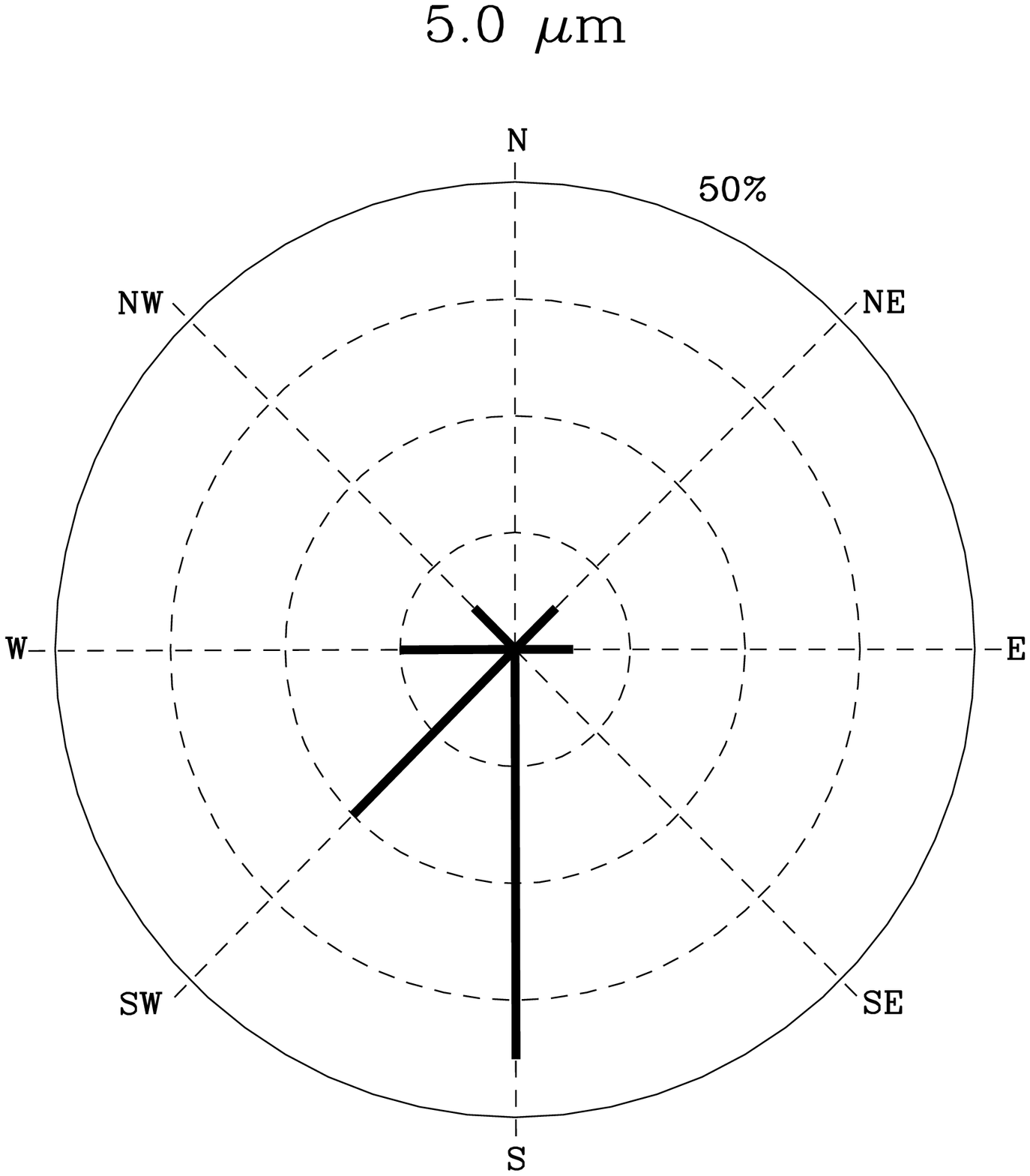}
\caption{Wind roses at TNG during dust storm events.}
\label{dust_vs_wdir}
\end{figure}
\begin{table}[b]
     \begin{center}
     \caption[]{Number of storm events at ORM in wintertime, summertime, and in the entire annual cycle in the period from August
2001 to December 2006.}
     \label{storm_seasonal}
     \begin{tabular}{c c c c}
     \hline\hline
 & wintertime & summertime & total\\
\hline
0.3 $\mu$m & 10 & 8 & 18\\
0.5 $\mu$m & 15 & 12 & 27\\
1.0 $\mu$m & 21 & 22 & 43\\
5.0 $\mu$m & 9 & 7 & 16\\
\hline
\hline
    \end{tabular}
    \end{center}
\end{table}
In a deeper analysis we noticed that  the dust storms are typically
3-4 days long, but sometimes it is possible to have some short
storms 2 days long or, more rarely, long-lasting storms persisting
5-6 days.
We correlated the wind direction with the dust-storm counts. Figure
\ref{dust_vs_wdir} shows the wind roses obtained during the storm
events as a function of particle size. The wind direction is
recorded by the TNG meteorological tower and data were already
analysed in Paper II. There is a clear correlation between dust
storms and wind direction. Only the winds blowing from south-west,
i.e. from the edge of the Caldera de Taburiente, carry  a large amount of dust particles. We confirm that
 the best photometric conditions occur when the wind blows from the north.\\
In Table \ref{storm_seasonal} the total number of events for both wintertime and summertime in the period from August 2001 to December
2006 is reported. Particles having 1.0 and 0.5 $\mu$m size show the larger number of events. Table \ref{storm_seasonal} shows that
dust storms occur with the same statistic in both wintertime and summertime.
Whittet et al. (\cite{whittet87}) demonstrated that $95\%$ of the
aeolian deposits collected at La Palma have diameters between 1 and
100 $\mu$m. To better understand this analysis we calculated the
median value of the dust content in case of storms for each measured
size, and
to include more massive grains we extrapolated the counts to 10.0
$\mu$m. Following Whittet et al. (\cite{whittet87}), we assume a
power-law distribution $r^{2}N(r)$ for particles $\geq$$1.0$ $\mu$m.
Because we cannot distinguish between seasons, we decided to
consider this background as a constant in both wintertime and
summertime. Under these assumptions we found that 10.0 $\mu$m
particles have a typical background of $\sim$$10$ N m$^{-3}$ which
is a lower concentration than the other particles we measured.
The same power law permits us to extrapolate the concentration to
the 10.0 $\mu$m size in case of dust storms. We obtained a
concentration of $\sim$$4\cdot 10^{4}$ N m$^{-3}$. Results of this
analysis are plotted in Figure
\ref{dust_distribution}.\\
\section{Aerosol atmospheric extinction in $B$, $V$, and $I$}\label{extinction}
\begin{figure}[t]
\centering
\includegraphics[width=8cm]{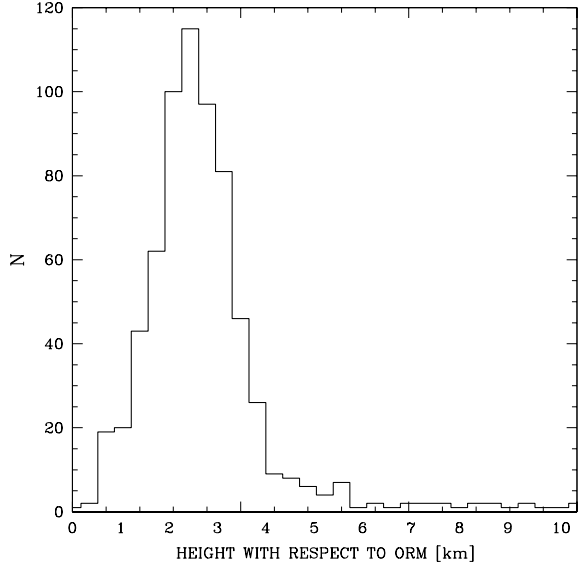}
\caption{Distribution of the dust altitudes from ORM altitude for
Eq. (\ref{eq:kaerVSkcamc}).} \label{dust_h}
\end{figure}

\begin{figure}[t]
\centering
\includegraphics[width=8cm]{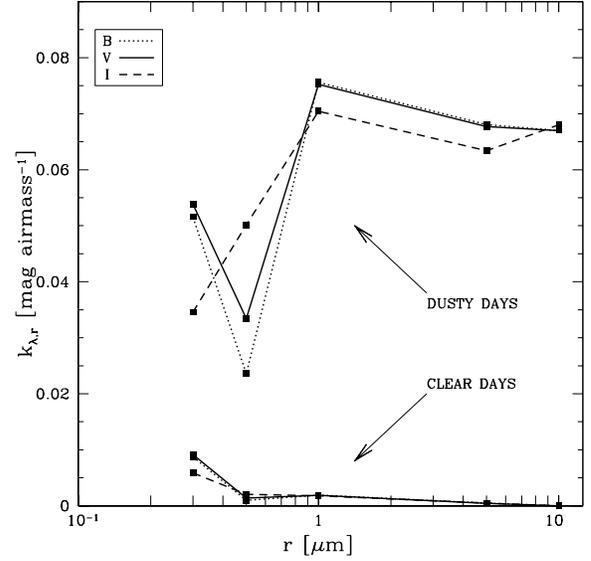}
\caption{Median aerosol atmospheric extinction in $B$, $V$, and $I$
in typical dust background conditions and in typical dust-storm
conditions for each particle size. The biggest particles are
dominant with respect to sub-micron particles.} \label{ext_avrg}
\end{figure}

\begin{figure}[t]
\centering
\includegraphics[width=8cm]{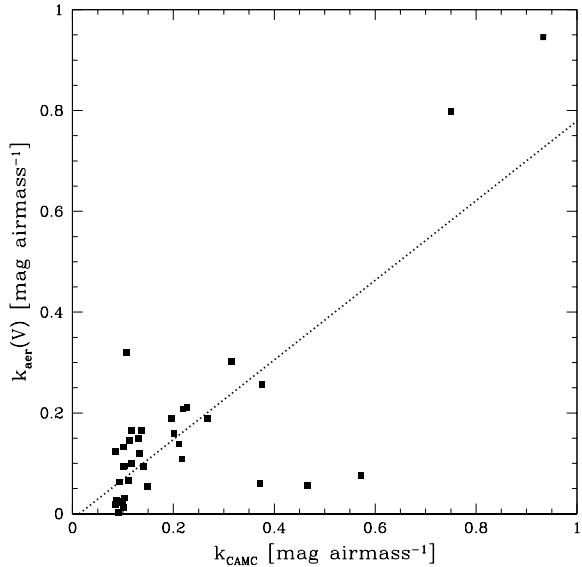}
\caption{$k_{aer}$ versus $k_{CAMC}$ in the case of dust storms. The
linear fit has a confidence level of 0.8.} \label{k_CAMC}
\end{figure}

\begin{figure}[b]
\centering
\includegraphics[width=8cm]{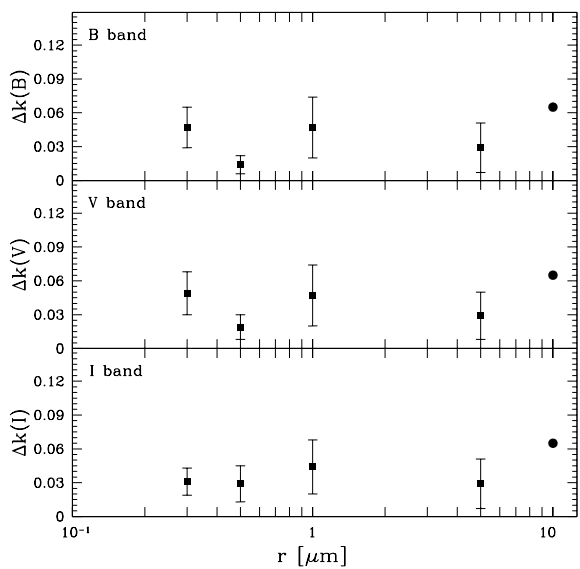}
\caption{Typical local aerosol extinction increases $\Delta k(\lambda)$ in [mag airmass$^{-1}$] in $B$ ($top$), $V$($center$) and $I$
($bottom$) in connection with dust storm events at ORM. The filled circles represent the expected $\Delta k(\lambda)$ due to 10 $\mu$m particles.}
\label{variations}
\end{figure}

The optical properties of the dust in La Palma have been
investigated by several authors who found that aerosol extinction
does not have a clear dependence on the adopted wavelength. In
particular, Stickland (\cite{stickland87}) pointed out that
extinction during dust storms is relatively stable on a time scale
of 15 minutes. No analysis exists for the effects of
 the different dust sizes on astronomical extinction.\\
In this paper we want to compare the differential extinction computed using our measured dust with simultaneous nighttime astronomical extinction
 obtained from the CAMC database, which is a very useful tool for quantitatively studying dust effects. We evaluate
aerosol atmospheric extinction ($k$) for each size of dust at 3
different wavelengths: 435, 550, and 780 nm corresponding to the
central wavelengths of the standard photometric bands $B$, $V$, and
$I$. We believe that using the differential dust counter may help us
evaluate the different extinction contributions from the different
dimensions of the grains as a function of the wavelength.\\
The aerosol extinction is obtained by applying the model based on
the Mie theory and has also been used by Mathis et al.
(\cite{mathis77}), Whittet et al. (\cite{whittet87}), Jim\'enez et
al. (\cite{jimenez98}), and Marley et al. (\cite{marley99}). The Mie
theory assumes the particles are homogeneous spheres of radius $r$.
Following Patterson (\cite{patterson77}) definitions, the scattering
coefficients for a Mie solution are linked to the size and
composition of the particles through the parameter $x = 2\pi
r/\lambda$. For an incident light at wavelength $\lambda$,  the
extinction coefficient $k$ due to particles of radius $r$ is defined
by the formula
\begin{equation}\label{eq:kappa}
\begin{array}{l}
k_{\lambda,r} = Q_{ext}(n,r,\lambda)\:\pi r^{2}\;N,
\end{array}
\end{equation}
where the term $Q_{ext}(n,r,\lambda)$ is the \textit{extinction
efficiency factor} computed for Mie's particles, while $N$ is the
number of particles of radius $r$ in a cubic centimetre. With this
definition,  $k_{\lambda,r}$ is expressed in magnitudes per
centimetre  [mag cm$^{-1}$] and is correlated with the \textit{local} density of the dust content in the atmosphere above the TNG dome.\\
In the equation, the extinction efficiency factor $Q_{ext}$ is a
complex function depending on the particles radius $r$, the absorbed
wavelength $\lambda$ and the complex refractive index $n$. In the
definition (Patterson \cite{patterson81}),  the real part of $n$ is
the ratio between light speed in a vacuum and the light speed in
particles, while the imaginary part is related to the Bouguer
absorption coefficient $\kappa$ through the formula $n_{im} =
\kappa\lambda/4\pi$. Following  Patterson (\cite{patterson77}),
Carlson \& Benjamin (\cite{carlson80}), Fouquart et al.
(\cite{fouquart87}),  and Jim\'enez et al. (\cite{jimenez98}),  we assigned
$n = 1.55 - 0.005i$ ($i = \sqrt{-1}$) as the typical refractive
index for Saharan dust aerosols.\\
\begin{figure}[t]
\centering
\includegraphics[width=8cm]{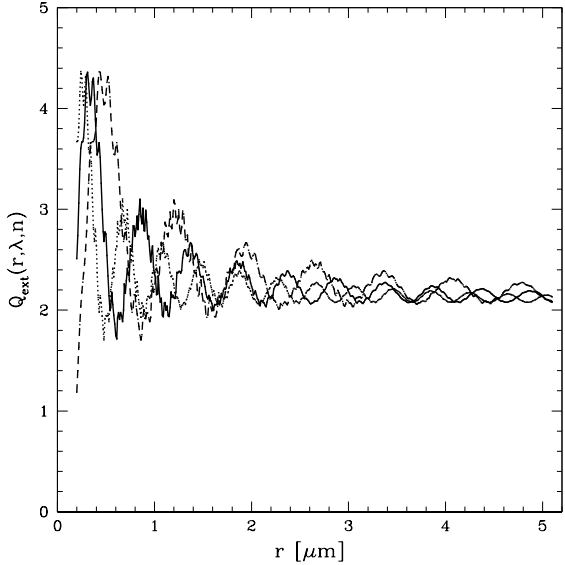}
\caption{Extinction efficiency factor as a function of the particles
radius in $B$ (dots), $V$ (solid), and $I$ (short-dashes).}
\label{Qext}
\end{figure}
\begin{table}[b]
     \begin{center}
     \caption[]{Values of the extinction efficiency factor $Q_{ext}$.}
     \label{Qexttab}
     \begin{tabular}{r c c c}
     \hline\hline
 & $B$ & $V$ & $I$\\
\hline
0.3 $\mu$m & 4.148 & 4.334 & 2.783\\
0.5 $\mu$m & 1.946 & 2.756 & 4.137\\
1.0 $\mu$m & 2.367 & 2.355 & 2.205\\
5.0 $\mu$m & 2.142 & 2.135 & 2.140\\
10.0 $\mu$m & 2.072 & 2.084 & 2.099\\
\hline
\hline
    \end{tabular}
    \end{center}
\end{table}
We computed the extinction efficiency factor $Q_{ext}$ for the
adopted photometric bands using the Mie scattering code of Wiscombe
(\cite{wiscombe80}). Figure \ref{Qext} shows the plot of $Q_{ext}$
as a function of the particle radius for $B$, $V$, and $I$ bands,
while in Table \ref{Qexttab} we report the computed values of
$Q_{ext}$ for 0.3, 0.5, 1.0, and 5.0 $\mu$m dust in the three bands.
We show that for $r \geq 1.0$ $\mu$m, $Q_{ext}$ is almost constant
in the three bands, confirming that for massive grains the
extinction is colour-neutral as pointed out by Jim\'enez et al. (\cite{jimenez98}).\\
Applying Eq. (\ref{eq:kappa}),  we obtained the local aerosol
atmospheric extinction in [mag cm$^{-1}$] caused by different grain
sizes in typical dust background conditions ($k_{MED}$) and in the
presence of dust storm events ($k_{dusty}$).\\
Table \ref{k_med} reports the $k_{MED}$ in $B$, $V$, and $I$  in
wintertime, summertime, and in the entire annual cycle using data as
reported in Table \ref{median_dust} and for the extrapolated value
of dust background at 10.0 $\mu$m. As shown in the table, on clear
days 0.3 $\mu$m dust gives the highest contribution to the global
extinction, in particular in summertime. The extinction due to 0.3
$\mu$m dust is lower at higher wavelengths, while for 0.5 $\mu$m
dust it becomes stronger at higher wavelengths, in particular in the
summertime. Finally, the effect of the grains starting from 5.0
$\mu$m  is almost constant in the three bands. We found that the
contribution of the extinction due to 10.0 $\mu$m particles is
negligible.\\
\begin{table}[t]
     \begin{center}
     \caption[]{Seasonal and annual local aerosol atmospheric extinction computed in dusty background
     conditions$^{\mathrm{\left[a\right]}}$.}
     \label{k_med}
     \begin{tabular}{r | c c c }
     \hline\hline
 & \multicolumn{3}{c}{wintertime}\\
 & $k_{MED}(B)$ & $k_{MED}(V)$ & $k_{MED}(I)$\\
\hline
0.3 $\mu$m & 1.49 & 1.56 & 0.99\\
0.5 $\mu$m & 0.19 & 0.26 & 0.39\\
1.0 $\mu$m & 0.38 & 0.38 & 0.35\\
5.0 $\mu$m & 0.11 & 0.11 & 0.11\\
10.0 $\mu$m & $< 0.01$ & $< 0.01$ & $< 0.01$\\
\hline
\hline
 & \multicolumn{3}{c}{summertime}\\
 & $k_{MED}(B)$ & $k_{MED}(V)$ & $k_{MED}(I)$\\
\hline
0.3 $\mu$m & 5.14 & 5.37 & 3.45\\
0.5 $\mu$m & 0.58 & 0.81 & 1.22\\
1.0 $\mu$m & 1.12 & 1.12 & 1.05\\
5.0 $\mu$m & 0.25 & 0.25 & 0.25\\
10.0 $\mu$m & $< 0.01$ & $< 0.01$ & $< 0.01$\\
\hline
\hline
 & \multicolumn{3}{c}{annual}\\
 & $k_{MED}(B)$ & $k_{MED}(V)$ & $k_{MED}(I)$\\
\hline
0.3 $\mu$m & 3.49 & 3.65 & 2.34\\
0.5 $\mu$m & 0.39 & 0.55 & 0.82\\
1.0 $\mu$m & 0.76 & 0.76 & 0.71\\
5.0 $\mu$m & 0.18 & 0.18 & 0.18\\
10.0 $\mu$m & $< 0.01$ & $< 0.01$ & $< 0.01$\\
\hline
\hline
    \end{tabular}
    \end{center}
         \begin{list}{}{}
\item{$^{\mathrm{\left[a\right]}}$ Values are given in [$10^{8}$ mag cm$^{-1}$].}
\end{list}
\end{table}
To understand the aerosol contribution on the total astronomical
extinction, we need to evaluate the distribution in altitude of the
particles. We compare our local extinctions in $V$ to simultaneous
astronomical extinctions in $V$ from the  CAMC database ($k_{CAMC}$)
under the assumption that $k_{CAMC}$ is mainly due to our measured
dust. We assume
\begin{equation}\label{eq:kaer}
\begin{array}{l}
k_{aer}(\lambda) = k_{\lambda,0.3} + k_{\lambda,0.5} + k_{\lambda,1.0} + k_{\lambda,5.0} + k_{\lambda,10.0}
\end{array}
\end{equation}
Because the CAMC-extinction database is only in $V$ band, we compare
our $V$ band aerosol extinctions. For each $k_{aer}(V)$ we assume
that particles are distributed in the atmosphere at an altitude $h$,
in order to have
\begin{equation}\label{eq:kaerVSkcamc}
\begin{array}{l}
k_{aer}(V)\:h = k_{CAMC}(V).
\end{array}
\end{equation}
We assume again that dust is uniformly distributed in altitude in a
column of atmosphere having a base of 1 cm$^{2}$ and height $h$ as
obtained from Eq. (\ref{eq:kaerVSkcamc}). Figure \ref{dust_h} shows
the distribution of the altitudes above the ORM obtained from each
computed mean dust count. The maximum of the histogram shows the
 altitude reached by the dust.
We found that the peak of the altitude distribution of the dust is
about 2.5 km above the telescope. This value agrees very closely
with studies based on independent radiosonde data (Hsu \cite{hsu99})
and indicates that the settlement process of the dust from the
Sahara to La Palma, in the size range 0.3-10.0 $\mu$m, is not
relevant, in agreement with the models of Murphy (\cite{murphy90}).
We conclude that, under the assumption of a uniform distribution of
the dust, the particles are distributed in the first 2.5 km, on
average. We then recompute  the aerosol atmospheric extinction using
Eq. (\ref{eq:kappa}), integrating on a column of atmosphere of 2.5
km.
We assumed this obtained altitude to be equivalent to 1 airmass and the final extinction is therefore given in [mag airmass$^{-1}$].\\
The $k_{\lambda,r}$ for each particle size in typical dust
background conditions and in typical dust storm conditions are now
recomputed by integrating on a 2.5 km column of atmosphere above the
TNG mirror.
Figure \ref{ext_avrg} shows the aerosol atmospheric extinction
$k_{\lambda,r}$ in [mag airmass$^{-1}$] and integrated on a column of atmosphere of
2.5 km in the entire annual cycle as a function
of the size of the dust (cfr. Table \ref{median_dust} and Figure
\ref{dust_distribution}) for clear and dusty days.\\
We show that on clear days, the extinction is dominated by 0.3
$\mu$m particles. Table \ref{dust_percentages} reports the
percentage of contribution by  different size of dust in each filter
on both clear and dusty days. As shown in the table, on clear days
0.3 $\mu$m particles are responsible for more than $70\%$ of the
total aerosol extinction in $B$ and $V$, while their contribution in
$I$ is about $60\%$. The biggest particles ($\geq$$1.0$ $\mu$m)
dominate in dusty days. It is interesting to note the different
behaviors of $I$ band during dust storms with respect to the $B$ and
$V$ bands. Bigger particles show a contribution to the total
extinction of more
than $70\%$ in all the bands (see Table \ref{dust_percentages}).\\
Following Eq. (\ref{eq:kaer}), the expected total local aerosol
atmospheric extinction has to be calculated as the sum of the
contributions of each particles size. On dusty days this extinction
is typically $\sim$$0.2$ mag airmass$^{-1}$ (see Figure
\ref{ext_avrg}) and can rise to $\sim$$1.0$
 mag airmass$^{-1}$ during very strong dust storms (see Figure
 \ref{k_CAMC}). Figure \ref{k_CAMC}
shows $k_{aer}(V)$ calculated on dusty days compared to simultaneous
astronomical extinctions in $V$ band as extracted from the CAMC
database. Figure \ref{k_CAMC} shows a good correlation, the dotted
line is the linear fit at a confidence level of 0.8 and having a
slope of 0.77. We can conclude that it is possible to have an
estimation of astronomical extinction, starting from a measure of
dust count and by  an appropriate calibration.\\
\begin{table}[b]
     \begin{center}
     \caption[]{Relative contribution of each particle size as percentage of the total aerosol atmospheric extinction
     $k_{aer}$.}
     \label{dust_percentages}
     \begin{tabular}{c | c c c c c}
     \hline\hline
 & \multicolumn{5}{c}{clear days}\\
 & 0.3 $\mu$m & 0.5 $\mu$m & 1.0 $\mu$m & 5.0 $\mu$m & 10.0 $\mu$m\\
\hline
$B$ & 72 &  8 & 16 & 4 & $< 0.1$\\
$V$ & 71 & 11 & 14 & 4 & $< 0.1$\\
$I$ & 58 & 20 & 17 & 5 & $< 0.1$\\
\hline
\hline
 & \multicolumn{5}{c}{dusty days}\\
 & 0.3 $\mu$m & 0.5 $\mu$m & 1.0 $\mu$m & 5.0 $\mu$m & 10.0 $\mu$m\\
\hline
$B$ & 18 & 8 & 26 & 24 & 24\\
$V$ & 18 & 11 & 25 & 23 & 23\\
$I$ & 12 & 17 & 24 & 23 & 24\\
\hline
\hline
    \end{tabular}
    \end{center}
\end{table}
To evaluate the increase in the local aerosol extinction in
connection with dust storm events, we also evaluated the differences
between (i) the aerosol extinctions calculated on each day of a dust
storm and (ii) the extinction calculated in the typical dust
background conditions of each defined seasonal epoch:
\begin{equation}\label{eq:Deltak}
\begin{array}{l}
\Delta k(\lambda) = k_{dusty}(\lambda)-k_{MED}(\lambda).
\end{array}
\end{equation}
Figure \ref{variations} plots the mean increases calculated for each
dust size in $B$, $V$, and $I$. In the figure we have also taken
 the extrapolated extinction due to 10.0 $\mu$m particles into account.
The typical increases are between 0.015 and 0.065 mag
airmass$^{-1}$.

\section{Summertime aerosol atmospheric extinction and TOMS aerosol index}\label{k_AI}
\begin{figure}[t]
\centering
\includegraphics[width=8cm]{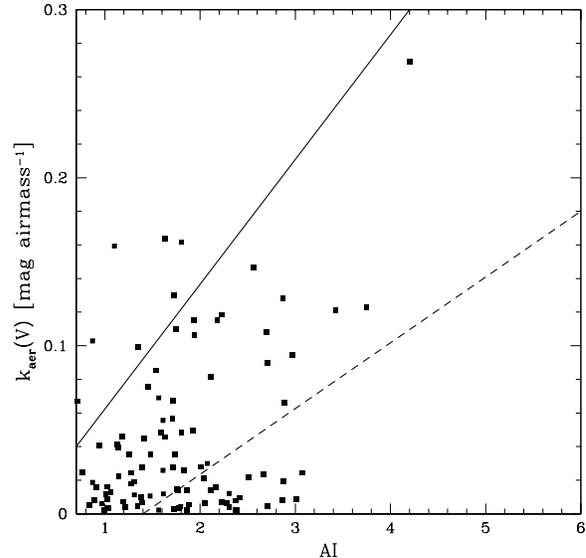}
\caption{Summertime local aerosol atmospheric extinction versus TOMS Aerosol Index in $V$ band.
The correlation calculated with the Spearman's test between the points inside the zone
delimited by the solid and the short-dashed lines has confidence level $> 0.8$.}
\label{k_VS_AI}
\end{figure}

Since from the extinction data alone it is not possible to
distinguish cirrus clouds from dust clouds, we have combined both
satellite and ground-based data comparing the daily
$k_{aer}(\lambda)$ computed in summertime in the three photometric
bands and the tropospheric TOMS aerosol index (AI) that
characterises the local aerosol concentration using satellite-based
reflectivity measurements. The TOMS data are set with  the data
closest to the ORM within a $1\times1$ deg latitude and longitude
box. TOMS AI is selected on the same days as $k_{aer}(\lambda)$.
Following Siher et al. (\cite{siher04}) we have set our threshold at
AI $>$ $0.7$. Because TOMS AI measurements above the Canaries are
done in daytime, we obtained the nighttime AI by interpolating the
values of two consecutive days. Satellite AI is obtained by the
reflectivity effect. Because both dust and clouds may show
reflectivity even at different threshold, we rejected AI points when
sky reflectivity $>$ $15\%$ (Bertolin \cite{bertolin05}) to be sure
to have only the dust effect.\\
Once more we split the data between wintertime and summertime, and
we noticed the absence of AI points in wintertime in connection with
our local aerosol extinction calculations. Most of those points are
the ones previously rejected for high sky reflectivity. In fact, in
wintertime the sky is often contaminated by high reflectivity caused
by thin cirrus or snow at the ground. Furthermore, in wintertime the
dusts are often at low altitudes, too deep in the atmosphere to be
revealed by the
satellite.\\
Figure \ref{k_VS_AI} shows the plot of the local aerosol atmospheric
extinction in summertime versus the AI in $V$ band ($B$ and $I$ show
similar behaviours). In the plot we can distinguish three areas. The
locus between the solid and short-dashed lines include all points
having a correlation between $k_{aer}(\lambda)$ and AI with a
confidence level $> 0.8$ (Spearman's test). The points below the
short-dashed line show high AI but low local extinction. This high
reflectivity is caused by the Saharan dust suspended at a very high
level that is  not detectable at the TNG mirror level. The (very
few) points above the solid line show high local extinction and low
AI. We ascribe
the cause to local, recycled dust that remain suspended at low altitude and cannot be revealed by the satellite.\\
To support this affirmation, we computed the two wind roses for the
data subsets relative to those days on which we believe that the
extinction is due to local recycled dust, and on those days when the
high reflectivity comes from the Saharan dust suspended in the high
atmosphere. We found that, in the case of local recycled dust, the
winds come prevalently from W-SW, while in the case of dust
suspended in the high atmosphere the prevailing direction is
from S-SW.\\
We can conclude that observations may often be affected also by
local recycled dust and not only by Calima events. More data may
help us this point. Taking  all the reasons mentioned above into
account we conclude that in site testing, both local dust monitors
and data from satellites have to be used together for a complete
investigation of the effects of the aerosols.

\section{Conclusions}\label{conclusions}
The high-altitude transport events of dust from desert areas have
been recognised the major atmospheric events on the Earth. The
resulting pollution has astronomical observations and
instrumentation (i.e. the main mirror reflectivity) effected even at very great distances,
 in particular at the Observatorio del Roque de Los Muchachos, on La Palma Island.\\
In this paper we have presented an analysis of the properties of the
local atmospheric dust measured at the height of the TNG primary
mirror, the correlations of the dust peak events with the wind
direction, and the total atmospheric extinction using the 5-year
long data archive of the TNG dust monitor. This particle counting
system is able to detect particles of 0.3, 0.5, 1.0, and 5.0 $\mu$m,
thereby according with the estimated dimension of dust and sand of
the  Sahara desert. Typical annual average background values are $3
\cdot 10^6$, $2.5 \cdot 10^5$, $10^5$, $10^3$ particles per cubic
metre, respectively, in the four channels. During dust storm events,
the particle concentration increases to three orders of magnitude.
Using a power-law distribution we also
extrapolated the contents of 10.0 $\mu$m dust on dusty days. We estimated an amount of $ 4 \cdot 10^4$ particles per cubic metre for massive grain.\\
After considering two seasonal epochs, we found the number of dusty
events to be comparable, but they reach a somewhat higher
concentration in summertime, when the background is also higher by a
factor 3-4. In both seasons the concentration of small dust
particles (0.3 $\mu$m) is
 dominant with respect to the other size dust particles.
 A study of dust-storm events versus wind direction has shown that winds coming from the southwest, i.e. from the edge
 of the Caldera de Taburiente, typically carry a large amount of dust particles.
 Dust storms are 3-4 days long, but sometimes it is possible to have
some shortly storms 2 days longer or, more rarely, long-lasting storms persisting 5-6 days.\\
A comparison between the storm duration and the TOMS Aerosol Index
shows that  short-storms typically are not linked with  a high
Aerosol Index. This suggests that short-storms can be generated by
local reprocessing of currents carrying dust in suspension at low
altitudes such is the TNG mirror level. We also noticed
  that dust storms occur not only in summertime (Saharan Calima events), but that several events
  occur during other seasons as effect of local currents recircle .\\
In clear days the aerosol extinction is
dominated by  0.3 $\mu$m particles ($\sim$$70\%$), while in dusty
days the strongest contribution is from bigger particles having size
$\geq$$1.0$ $\mu$m ($>$$70\%$). On dusty days the  extinction is
typically $\sim$$0.2$ mag airmass$^{-1}$, and it can rise to
 $\sim$$1.0$ mag airmass$^{-1}$ in the presence of very strong dust storms.\\
In dusty days dust $V$ extinction shows a positive correlation with simultaneous astronomical extinctions values from  CAMC database (c.l. 0.8).
This suggests that the dust detected at TNG affects the astronomical observations
in a very wide area.\\
Assuming a uniform distribution of the dust in the observing column,
we found that the dust extends about 2.5 km  above the telescope,
corresponding to an height of 5 km above the see level. This confirms that the settlement process of the dust
in the size range 0.3-10.0 $\mu$m, is not relevant.\\

\begin{acknowledgements}
The authors acknowledge the anonymous reviewer for the very useful comments and the TNG staff for the availability of the data.
The authors thank also Guido Barbaro of Department of Astronomy of University
of Padova for helpful comments on Mie's theory and Chiara Bertolin of Atmospheric and Climate Science Institute of
  National Research Council of Padova for the help in using TOMS/EarthProbe data.
\end{acknowledgements}

\end{document}